\documentclass[journal,transmag]{IEEEtran}
\usepackage{cite}
\ifCLASSINFOpdf
   \usepackage[pdftex]{graphicx}
   \graphicspath{{Images/}}
\fi
\usepackage{amsmath}
\usepackage[T1]{fontenc}
\usepackage{lipsum}
\usepackage{array}
\usepackage{algorithm}
\usepackage[noend]{algpseudocode}
 \usepackage[utf8]{inputenc}
\begin{document}
\title{On the utility of Power Spectral Techniques with feature selection Techniques for Effective Mental Task Classification in Non-invasive BCI}
\author{\IEEEauthorblockN{Akshansh Gupta\IEEEauthorrefmark{1},
Ramesh Kumar Agrawal\IEEEauthorrefmark{1},
Jyoti Singh Kirar\IEEEauthorrefmark{1}, 
Javier Andreu-Perez\IEEEauthorrefmark{2,3},\\
Wei-Ping Ding\IEEEauthorrefmark{4},
and Chin-Teng Lin\IEEEauthorrefmark{5},Mukesh Prasad\IEEEauthorrefmark{5}
}
\IEEEauthorblockA{\IEEEauthorrefmark{1}School of Computer and Systems Sciences, Jawaharlal Nehru University, New Delhi, India}

\IEEEauthorblockA{\IEEEauthorrefmark{2}School of Computer Science and Electronic Engineering, University of Essex, United Kingdom}
\IEEEauthorblockA{\IEEEauthorrefmark{3}Faculty of Medicine, Imperial College London United Kingdom}
\IEEEauthorblockA{\IEEEauthorrefmark{4}School of Computer and Technology, Nantong University, Nantong, China}
\IEEEauthorblockA{\IEEEauthorrefmark{5}Centre for Artifical Intelligence, School of Software, FEIT, University of Technology Sydney, Australia}
}
\IEEEtitleabstractindextext{%
\begin{abstract}
In this paper classification of mental task-root Brain-Computer Interfaces (BCI) is being investigated, as those are a dominant area of investigations in BCI and are of utmost interest as these systems can be augmented life of people having severe disabilities. The BCI model's performance is primarily dependent on the size of the feature vector, which is obtained through multiple channels. In the case of mental task classification, the availability of training samples to features are minimal. Very often, feature selection is used to increase the ratio for the mental task classification by getting rid of irrelevant and superfluous features. This paper proposes an approach to select relevant and non-redundant spectral features for the mental task classification. This can be done by using four very known multivariate feature selection methods viz, Bhattacharya's Distance, Ratio of Scatter Matrices, Linear Regression and Minimum Redundancy \& Maximum Relevance. This work also deals with a comparative analysis of multivariate and univariate feature selection for mental task classification. After applying the above-stated method, the findings demonstrate substantial improvements in the performance of the learning model for mental task classification. Moreover, the efficacy of the proposed approach is endorsed by carrying out a robust ranking algorithm and Friedman's statistical test for finding the best combinations and comparing different combinations of power spectral density and feature selection methods.
\end{abstract}
}
\maketitle
\IEEEdisplaynontitleabstractindextext
\IEEEpeerreviewmaketitle
\section{Introduction} \label{Intro} A Brain-Computer Interfaces (BCI) \cite{graimann2010brain,rahman2018efficient} is a message transmission framework, through which an individual can communicate for necessities by his or her brain signals, even absence of normal pathway of the computer system and a very effective device for the person with severe motor impairment \cite{gandhi2014eeg,zhao2016ssvep}.  It is pragmatic area, which has focused to the design and invent of neuron rooted means to endue solutions for disease prediction , communication and control \cite{anderson1998multivariate}, \cite{babiloni2000linear}, \cite{keirn1990new}. On the ground of acquisition of the brain signal BCI is broadly divided in three categories in literature \cite{kubler2000brain}, \cite{schalk2008brain}, viz,  invasive, semi-invasive (electrocorticography (ECoG)) and non-invasive(electroencephalography EEG). Economically nature \cite{hewireless} and calibre to capture brain signals in a non-invasive fashion, EEG is a mostly preferred technique to aquire brain activity for BCI systems \cite{akram2014p300}, \cite{keirn1990new}. BCI systems can be used as a \textit{Response to mental tasks} system, \cite{bashashati2007survey}, which is perceived to be more practical for locomotive patients. The basic assumption of this type of system is that mental activities lead to produce task-originated patterns. The BCI system's success depends on the precision of classification assorted mental tasks. These tasks requires extractions of discriminative features from the raw EEG signal to distinguish different mental tasks \cite{zhang2017classification}.\\
In previous studies, the researchers have utilized plenty approaches of feature extraction for better representation of the EEG signal for the classification process in the BCI domain, for example Band Power \cite{pfurtscheller1997eeg}, amplitude values of EEG signals \cite{kaper2004bci}, Power Spectral Density (PSD)\cite{chiappa2004hmm,neshov2018classification}, \cite{moore2003real}, \cite{palaniappan2002new}, Autoregressive (AR) and Adaptive Autoregressive (AAR) parameters \cite{penny2000eeg}, time-frequency and inverse model-based features \cite{qin2004motor}, \cite{kamousi2005classification},\cite{congedo2006classification}. Wavelet Transform (WT) \cite{mallat1989theory},\cite{gupta2012three} and Empirical Mode Decomposition (EMD) \cite{huang1998empirical}, \cite{diez2009classification},\cite{gupta2012relevant},\cite{bajaj2012classification},\cite{rutkowski2010emd}, \cite{fine2010assessing},\cite{mylonas2016modular} have been used to decompose non-stationary and non-linear EEG signals into smaller frequency components. However, both WT and EMD methods provide low-frequency resolution and may not handle efficiently different overlapping frequency bands \cite{battista2007application}, \cite{adamczak2010investigating}  present in the EEG. On the other hand, power spectral analysis provides high-frequency resolution. The recording of EEG data occurs from multiple sensors/channels. Hence, the EEG data contains huge number of features but the recording session of the person usually very small in number. That produces, a small number of data samples.  Hence, it suffers the curse of dimensionality as the ratio of features and sample is very small \cite{bellman1961adaptive}. To overcome this problem, reduction of the dimension using feature selection is suggested in literature \cite{guyon2003introduction}.  In spite of that, no in-depth study has ever been conducted about how to  use power spectral features effectively with combination feature selection techniques in BCI the applications.\\
In this article we provide answers to the following questions:
\begin{enumerate}
\item Whether extraction of features using power spectral techniques helps in mental task classification.
\item Whether the further reduction in dimensionality of features using feature selection approaches improves the classification performance or not.
 \item  Is multivariate feature selection approach better than univariate feature selection approach?
\item  Which conjunction of feature extraction and selection method performs best for mental task classification? 
\end{enumerate}
Thus, this present work proposes a procedure of the determination of a compact collection of \textit{relevant and non-redundant features} from the EEG signal in the two-phase approach. The first phase elaborates about the extraction of  PSD features from the EEG signal using three different approaches. In the second level, a set of relevant and non-redundant features is sorted by multivariate filter feature selection approach. 
 To investigate the performance of different combinations of PSD method and multivariate feature selection method, experiments are conducted on an open EEG data \cite{keirn1990new} source. The performance is calculated in terms of classification accuracy and compared with a combination of PSD and a univariate filter feature selection method. 
 In order to rank and compare multiple combinations of power spectral density and feature selection methods Ranking method and Friedman's statistical test were also performed. \\

The rest of the paper is organized as follows: The Power Spectral Estimation approach has been discussed briefly in Section ~\ref{PSD}.The proposed approach to obtain minimal subset of relevant and non-redundant of the PSD features using multivariate feature selection methods is included in the Section ~\ref{PA}. The Descriptive information, data and method results are presented in Section ~\ref{EX}. In the final, Section ~\ref{CON} conclusions and future work is discussed.
\section{Feature Extraction using Power Spectral Density}\label{PSD} Power Spectral Density (PSD) is a measure of average power associated with any random sequence \cite{stoica2005spectral}, which can be catalogued into three categories: (i) Non-parametric, (ii) Parametric and (iii) Subspace. The non-parametric methods are simple to compute and robust. Periodogram based estimation, Bartlett Window, Welch window and Blackman and Tuckey method are examples of this category. However, they do not provide the necessary frequency resolution due to their inability to extrapolate the finite length sequence for data points exceeding the signal length. Another drawback of this approach is spectral leakage \cite{proakis2001digital}. To overcome the drawback of non-parametric methods, parametric method is suggested. The estimation of PSDs values from a given signal in parametric approaches are carried out by assuming that output of the linear system is driven by white noise and then parameters of the system are calculated. Examples are the \textit{Yule-Walker autoregressive }(\textit{AR}) \textit{method} \cite{pfurtscheller1998separability}, the \textit{Burg method} \cite{chiappa2004hmm}, Covariance and modified covariance etc. The commonly used parametric linear system model is the all-pole model which consists of a filter with all zeroes at the origin and occurs in the z-plane. The output produced by such a filter using white noise as input is an autoregressive (AR) process. Thus, these spectral estimation methods are also sometimes known as \textit{AR methods}. The AR methods tend to aptly describe data spectrum that is "peaky", the data having PSDs value large at certain frequencies, e.g. speech data. Smoother estimates of the PSD are produced by parametric methods than non-parametric methods but are subject to error if the order of model is not chosen correctly. Sub-Space methods are often used when SNR is low. PSDs values are obtained concerning  Eigen-decomposition of autocorrelation matrix. For line spectra or spectra having sinusoidal nature Subspace methods are better choice and are also effective in the recognition of sinusoids mixed in noise.  However, the subspace methods suffer from the following:The method in all probability does  not generate true PSD estimates; it does not store power which is required for processing between the time and frequency domains; and it flunks  in getting back the autocorrelation series by computing the inverse Fourier transform of the frequency estimate.\\
For a given stationary random signal $\mathbf{x}_m $, the PSD $P_{xx}$ is mathematically related to the autocorrelation sequence by  Fourier transform, which regarding normalized frequency $f_{s}$, is given by,
\begin{equation} 
{\ P}_{xx}\left(f\right)=\frac{1}{f_s}\sum^{\infty }_{m=-\infty }{R_{xx}}\left(m\right)e^{-\frac{j2\pi mf}{f_s}} 
\end{equation}  
where $f_s$ is the sampling frequency. Fourier transform of the autocorrelation of the signal also gives the PSD. Using the inverse discrete-time Fourier transform from the PSD the correlation sequence can be derived as following:
\begin{equation} 
R_{xx}=\int^{\pi }_{-\pi }{P_{xx}}\left(\omega \right)e^{-j\omega m}d\omega =\int^{{f_s}/{2}}_{{-f_s}/{2}}{P_{xx}}\left(f\right)e^{{j2\pi f}/{f_s}}df
\end{equation} 
The average power of the sequence \textit{x}${}_{n}$ over the entire Nyquist interval is represented by
\begin{equation}  
R_{xx}\left(0\right)=\int^{\pi }_{-\pi }{P_{xx}}\left(\omega \right)d\omega =\int^{{f_s}/{2}}_{{-f_s}/{2}}{P_{xx}}\left(f\right)df
\end{equation} 
For a particular frequency band [$\omega $${}_{1}$, $\omega $${}_{2}$], $\left( 0\le {{\omega }_{1}}\le {{\omega }_{2}}\le \pi  \right)$, the average power of a signal is given by:
\begin{equation} \
\overline{P_{\left[{\omega }_1,{\omega }_1\right]}}=\int^{{\omega }_2}_{{\omega }_1}{P_{xx}}\left(\omega \right)d\omega   
\end{equation} 
It can be seen from the above expression that ${{\rm P}}_{{\rm xx}}\left({\rm w}\right)$ represents the power content of a signal in an \textit{extremely small} frequency band, which is why it is known as the power spectral \textit{density}. 
\subsection{Welch Method}
This method falls under non-parametric approach. For a finite time duration random signal $\mathbf{x}_m$ of $N$ interval length, PSD values are estimated with the help of a periodogram which is the squared modulus of discrete Fourier transform of the signal and is given by
\begin{equation} 
{\ P}_{\mathbf{x}\mathbf{x}}\left(f\right)=\ \frac{1}{N}{\left|\mathbf{x}\left(f\right)\right|}^2  
\end{equation} 
Here $f$ corresponds to the frequency of the sequence and $X(f)$ is the Fourier transform of the signal. A periodogram gives asymptotically non biased estimate of power spectrum. 

In Welch method, $N$ length signal is divided into $K$ overlapped segments each of length $M$.  The $i^{th}\ $segment is given by,
\begin{equation} 
{{\rm \ \ \ \ \mathbf{x}}}_{{\rm i}}\left({\rm n}\right){\rm =\mathbf{x}}\left({\rm n+iD}\right)
\end{equation} 
Here $n=0$\dots $N-1$, $i=0$\dots $K-1$ and $D$ is overlap segment. For this, a windowed segment periodogram is given by                                                                                                            \begin{equation}
{ {P}^i_{XX}\left(f\right)=\frac{1}{MU}{\left|\sum^{N-1}_{i=0}{w(n){{\mathbf x}}_i(n)e^{-j2\pi fn}}\right|}^2\ }
\end{equation}

where $w(n)$ is the window function and $U$ is the power of the window function given by, 
\begin{equation}  
{ U=\frac{1}{M}\sum^{M-1}_{n=0}{w^2\left(n\right)}} 
\end{equation} 

The average of $K$ periodograms depicts Welch power spectrum and is given by: 
\begin{equation} 
{{P}^W_{XX}=\frac{1}{K}\sum^{K-1}_{i=0}{P^i_{XX}\left(f\right) }} 
\end{equation} .
\subsection{Burg Method}
The Burg method \cite{stoica2005spectral} is a parametric method of spectral analysis. The PSDs values can be obtained by finding $pth$ order coefficients of an AR process. A $pth$ order real valued AR signal $\mathbf{x}(n)$ (with zero mean) at point $n$ is given by \cite{palaniappan2002new}.
\begin{equation} 
\mathbf{x}\left(n\right)=-\sum^p_{m=1}{a_mx\left(n-m\right)}+e(n)
\end{equation} 
Here $a_m$ is AR coefficient of $x(n-m)$, $e(n)$ is the error term at point $n$ independent of past terms. Burg algorithm test to find the AR coefficient by applying more data points and minimizes the forward and backward prediction errors in the least squares sense \cite{palaniappan2002new}, with the AR coefficients constrained to satisfy the Levinson-Durbin recursion. It provides high resolution for short data records. 
After finding AR coefficients by Burg Algorithm, PSD value $S(f)$ at frequency $f$ is given by:
\begin{equation}  
\ S\left(f\right)=\frac{S_e(f)}{{\left|1+\ \sum^p_{i=1}{a_ie^{-j2\pi fiT}}\right|}^2} 
\end{equation} 
Here $T$ is the sampling period and $S_e\left(f\right)$ is spectrum of error sequence which should be flat i.e. independent of frequency. One of foremost concern in AR modelling is the choice of order \textit{p. }To determine $p$, several criterion such as final prediction error (FPE)\cite{akaike1969fitting}, minimum description length \cite{rissanen1983universal}, Akaike information criterion (AIC)\cite{akaike1974new}, and   autoregressive transfer function \cite{parzen1957consistent} are proposed in literature. Among these, AIC is commonly used, which is given by
\begin{equation}  
 AIC\left(p\right)=ln{\sigma }^2_{wp}+\frac{2p}{n} 
\end{equation} 
where${\ \ \sigma }^2_{wp}$ is estimated variance in linear prediction error. From Table ~ \ref{Table-1}, it can be observed that AIC value is minimum for order 5 or 6. We have chosen p=6 in our experiments which is also suggested by Kerin \& Aunon \cite{keirn1990new}.
\begin{center}
\begin{table}
\caption{Variation of AIC value for a given order and a mental task.}
\label{Table-1}
\centering
\begin{tabular}{|l|l|l|l|l|}
\hline
Task     & \multicolumn{4}{c|}{Order}            \\ \hline
         & 5       & 6       & 7       & 8       \\ \hline
Baseline & -1.012  & -1.0117 & -1.0109 & -1.0106 \\ \hline
Count    & -1.2841 & -1.2851 & -1.2847 & -1.2842 \\ \hline
Letter   & -1.2574 & -1.259  & -1.2589 & -1.2585 \\ \hline
Math     & -1.2783 & -1.2772 & -1.2762 & -1.2768 \\ \hline
Rot      & -1.177  & -1.1768 & -1.176  & -1.1758 \\ \hline
\end{tabular}
\end{table}
\end{center}
\subsection{Multiple Signal Classification (MUSIC)}
 Music is an orthogonal subspace decomposition method is based on Pisarenko idea \cite{kia2007high} that allows the estimation of low Signal-to-Noise ratio (SNR) frequency components. This method is used to lowers the effect of noise in the analysed signal and finds the optimal frequency resolution in a dynamic signal \cite{ubeyli2008implementing}. Subspace method assumes that any discrete time signal $s[n]$ is representable in the form of $m$ complex sinusoids with a noise $p[n]$ such that
\begin{equation} 
s\left[n\right]=\sum^m_{i=1}{\overline{A_i}\ e^{j2\pi f_i}}+p\left[n\right],\ n=0,1,2,\dots ,N-1 
\end{equation} 
where  $\overline{A_i}=\left|A_i\right|e^{{\emptyset }_i}$ is magnitude of $i^{th}$ complex sinusoid, $m,N,f_i\ and\ {\emptyset }_i$ are frequency signal dimension order, number of sample data, frequency and phase of $i^{th}$ complex sinusoid.

The autocorrelation matrix $\mathbf{R}$ of signal $s[n]$ is given by:
\begin{equation}  
{\mathbf R}=\sum^m_{i=1}{{\left|A_i\right|}^2p\left(f_i\right)p^H\left(f_i\right)+\ {\sigma }^2{\mathbf I}} 
\end{equation} 
where $p\left(f_i\right)={\left[1\ e^{j2\pi f_i}\ e^{j4\pi f_i\ }\dots \ e^{2\pi {\left(N-1\right)f}_i}\right]}^T$  and  ${\sigma }^2$ is variance of white noise signal, H is hermitian transpose and I is the identity matrix.
 Therefore, it can be observed that $\mathbf{R}$ is a composition of sum of signal and noise autocorrelation matrices such that
\begin{equation}  
{\mathbf R}{\mathbf =}{{\mathbf R}}_{{\mathbf s}}{\mathbf +\ }{\sigma }^2{\mathbf I} 
\end{equation} 
Pisarenko has noticed that variance of noise acts with the smallest eigenvalues of $\mathbf{R}$. The  orthogonality of the signal and noise subspace is given as
\begin{equation}
p{{({{f}_{i}})}^{H}}v(m+1)=0,i=1,2,...,m
\end{equation}
where $v(m+1)$ is the eigenvector of noise in matrix $\mathbf{R}$ with dimension of $(m+1)\times(m+1)$
The estimation of PSD by Pisarnako is defined as
\begin{equation}
{{P}_{Pisarnako}}=\frac{1}{{{\left| p{{\left( {{f}_{i}} \right)}^{H}}v\left( m+1 \right) \right|}^{2}}}
\end{equation}
PSD estimation by MUSIC gives better performance than Pisarenko due to addition of averaging of extra noise eigenvectors$(k=m+1,m+2,~\ldots ,M)$.
Estimation of PSDs by MUSIC is given by:
\begin{equation}  
P_{MUSIC}\left(f\right)=\frac{1}{\sum^M_{k=m+1}{{\left|{p(f)}^Hv_{(k)}\right|}^2}} 
\end{equation} 
Here $~p{{(f)}^{H}}{{v}_{(k)}}=0$ for $k=1,\ldots ,m$  using orthogonality of the signal and noise subspace. These PSD values have major peaks at the principal components only. The performance of Music depends on the dimension of the autocorrelation matrix $(M\le N)$
\section{Proposed Approach-Feature Selection}\label{PA}
The number of PSD values obtained using one of the given three methods from multiple channels would be large, otherwise the number of training samples available is in general relatively small. Hence the method suffers from curse-dimensionality problem \cite{bellman1961adaptive}
.In order to subdue this problem, there is a need to determine a minimal set of pertinent features which can improve classification accuracy of a learning system. This work has proposed an approach to find a minimal subset of relevant feature using multivariate feature selection methods.

Feature selection \cite{kohavi1997wrappers}, \cite{guyon2003introduction} is one of the widely accepted approach to determine relevant features. In spite of available plenty of research works for the feature selection, not much work has been carried out in the domain of mental task classification. The filter and the wrapper approaches are the two major approaches of feature selection techniques. In filter approach, the step of selecting optimal features set is considered as one of the pre-processing steps of just before applying any machine learning algorithm. This approach adopts only inherent properties of the features and does not consider any virtue of any learning algorithm. Hence, it may not select the optimal feature set for the learning algorithm. Instead, the wrapper approach \cite{kohavi1997wrappers} finds an optimal features subset, which is compatible with the given learning algorithm. The given classifier requires to be trained for each feature of set of the all features separately in the wrapper approach, which makes it more computational costly than filter approach. \\
Filter approach is further partitioned in two categories on the basis of  the way of opting features \cite{guyon2003introduction}, as Univariate (single feature ranking) and Multivariate (feature subset ranking). Univariate method utilizes a scoring function for measuring relevance of the feature. Implementation of the Univariate method is very simple. In BCI field the researchers, \cite{koprinska2010feature},\cite{rodriguez2013efficient},\cite{guerrero2010eeg},\cite{murugappan2010classification} used univariate filter method. The performance of learning model usually improve with the help of reduced set of relevant features obtained by Univariate feature selection method. But it does not captures the correlation among the features. Hence there may be many redundant features in the subset of relevant feature which may take down the performance of learning model. Wrapper method, \cite{bhattacharyya2014automatic}, \cite{dias2010feature}, \cite{keirn1990new} has been applied to obtain a subset of  non-redundant features for the mental task classification. Due to high- dimensionality of feature of EEG data, wrapper approach is not feasible option for mental task classification as it will become more computationally expensive. Hence we have applied both uni-variate as well as multivariate filter feature selection algorithms.\\
Let us assume we have a data matrix $\mathbf{X}$, of $m$ rows,and $k+1$ columns, with data sample $\mathbf{x}_{i}, i=1,\ 2,\ \ldots, m$; containing features set $S=\mathbf{f}_{1}, \mathbf{f}_{2}, \ldots \mathbf{f}_{k}$ and class label $C_{1}, C_{2}, \ldots C_{n}, \text{where}\ n\leq m$.
\subsection{Uni-variate Feature selection}
\subsubsection{Pearson's Correlation}
Pearson's correlation coefficient (CORR), \cite{pearson1920notes,dowdy2011statistics} is employed to determine linear relationship between two variable. CORR of \textit{i}$^{th}$ feature vector (\textbf{f}$_i$) with the class label vector (\textbf{c}) is given by
 \begin{eqnarray}
CORR\left( {{\mathbf{f}}_{i}},\mathbf{c} \right)=\frac{cov({{\mathbf{f}}_{i}},\mathbf{c})}{{{\sigma }_{{{\mathbf{f}}_{i}}}}{{\sigma }_{c}}}=\frac{E[({{\mathbf{f}}_{i}}-{{\mu }_{i}})(\mathbf{c}-\bar{c})}{{{\sigma }_{{{\mathbf{f}}_{i}}}}{{\sigma }_{c}}}
 \end{eqnarray}
where $ i=1,\ 2,\ \ldots ,\ k$,  ${\sigma }_{{{\mathbf f}}_{i}}{,\ \sigma }_{{\mathbf c}}$ represent respectively the standard deviations of feature vector  ${{\mathbf f}}_{i}$ and ${\mathbf c}$. $cov\left({{\mathbf f}}_{i}{\mathbf ,\ }{\mathbf c}\right)$ represents the covariance between  ${{\mathbf f}}_{i}$ and ${\mathbf c}$, ${{\mu }_{i}}=\frac{1}{k}\underset{i=1}{\overset{k}{\mathop \sum }}\,{{X}_{ik}}$  and $\bar{c}=\frac{1}{k}\underset{i=1}{\overset{k}{\mathop \sum }}\,{{c}_{i}}$  are the mean of \textbf{f}$_k$ and \textbf{c} respectively.

Range of  $CORR\left({{\mathbf f}}_{i},{\mathbf c}\right)$ falls between -1 \& +1. The value nearby to $|1|$, depicts the stronger linear relation among the prescribed variables while zero value implies no correlation between the two variables.
\subsubsection{Mutual Information}
Mutual information [MI] is a feature ranking method on basis of Shannon entropy, which determines relationship between two variables. MI of a feature vector \textbf{f}$_i$ and the class vector $\mathbf{c}$ can be calculated as\cite{shannon1949mathematical}:
\begin{equation}
I({\mathbf{f_\mathit{i}}},\mathbf{c})=\sum P({\mathbf{f_\mathit{i}}},\mathbf{c})\log\frac{P({\mathbf{f_\mathit{i}}},\mathbf{c})}{P({\mathbf{f_\mathit{i}}})P(\mathbf{c})}
\end{equation}
 where $P({\mathbf{f_\mathit{i}}})$ and $P(\mathbf{c})$ are the marginal probability distribution functions for random variables ${{\mathbf f}}_{{\mathit i}}$ and \textbf{c} respectively and $P({\mathbf{f_\mathit{i}}},\mathbf{c})$ is joint probability distribution. The most extreme estimation of MI demonstrates the higher reliance of the variable on the class label. The advantage of MI is that it can discover even the non-linear dependency between the attribute and the relating class label vector \textbf{c}. 
 
\subsubsection{Fisher Discriminant Ratio }
Fisher Discriminant Ratio (FDR) is a univariate filter feature selection technique which depends on the statistical virtue of the attributes or features. FDR (${{\mathbf f}}_{{\mathit i}}$) for $i^{th}$ features for two class $C_{1}$ and $C_{2}$ can be given as:
\begin{equation}
FDR({{\mathbf{f}}_{i}})=\frac{{{({{\mu }_{1(i)}}-{{\mu }_{2(i)}})}^{^{2}}}}{\sigma _{1(i)}^{2}+\sigma _{2(i)}^{2}}
\end{equation}
here, ${\mu}_{ 1(i)}$ and ${\sigma}^{2}_{1(i)}$ are the mean and deviation of the data of class $C_{1}$ respectively for $i^{th}$ feature. 
\subsubsection{Wilcoxon's Ranksum Test}
Wilcoxon Ranksum Test, suggested by \cite{wilcoxon1945individual}, is a non-parametric statistical test, accomplishes  between data of two classes on the basis of median of the samples having no prior knowledge of probability distribution.

 The statistical distinctness $t(\mathbf{f}_i)$ of feature $\mathbf{f}_i$ for known two classes, class $C_{1}$ and $C_{2}$ using Wilcoxon's statistics can be defined as \cite{li2008gene}:
\begin{equation}
t\left( {{\mathbf{f}}_{i}} \right)=\underset{l=1}{\overset{{{N}_{i}}}{\mathop \sum }}\,\underset{m=1}{\overset{{{N}_{j}}}{\mathop \sum }}\,DF(({{X}_{li}}-{{X}_{mi}})\le 0)
\end{equation}
where $N_i$ and $N_j$ are the number of the data example in class $C_{1}$ and $C_{2}$  respectively, $DF$ is the logical discriminative mapping between two classes of data, which defines an estimation of 1 or 0 corresponding to true or false and $X_{li}$, is the expression values of $i^{th}$ feature for $l^{th}$ sample. The value of $t(\mathbf{f}_i)$ lies between zero and $(N_i\times N_j)$. The relevance of the feature can be fined as:
\begin{equation}
R\left( t({{\mathbf{f}}_{i}}) \right)=\text{max}(t({{\mathbf{f}}_{i}}),{{N}_{i}}\times {{N}_{j}}-t({{\mathbf{f}}_{i}}))
\end{equation}
\subsection{Multivariate Feature Selection}
Time-efficient multivariate filter method picks  a subset of features, which are relevant to the class label of data and independent from each other. Thus it up dues the limitations of both uni-variate and wrapper approaches. Thus we have opted most widely utilized multivariate filter methods by research community for the dimensionality reduction, are Bhattacharya distance measure \cite{bhattacharyya1946measure}, Ratio of scatter matrices \cite{devijver1982pattern}, Linear regression \cite{park2007forward} and minimum Redundancy-Maximum Relevance (mRMR) \cite{peng2005feature}. A brief discussion on the mentioned techniques is given below.
\subsubsection{Bhattacharaya Distance}
In literature, Bhattacharya distance is used for find similarity between two continuous or discrete probability distribution. It is a special case of Chernoff distance that provides similarity overlap of the distribution. For multivariate normal probability distribution, Chernoff Distance measure can be written as \cite{chernoff1952measure}:
\begin{equation}
\begin{split}
J_c=\frac{1}{2}\beta(1-\beta)(\boldsymbol{\mu}_{2}-\boldsymbol{\mu}_{1})^{T}[(1-\beta)\boldsymbol{\Sigma}_{1}+\beta\boldsymbol{\Sigma}_{2}]^{-1}(\boldsymbol{\mu}_{2}-\boldsymbol{\mu}_{1})+\\ \frac{1}{2}log\frac{\left | (1-\beta)\boldsymbol{\Sigma}_{1}+\beta\mathbf{\Sigma}_{2} \right |}{\left |\mathbf{\Sigma}_{1}  \right |^{1-\beta}\left |\mathbf{\Sigma}_{2}  \right |^{\beta}}
\end{split}
\end{equation}
where ${\boldsymbol{\mu}}_{i}$ and $\mathbf{\Sigma}_{i}$ are mean vector and covariance matrix for class $C_{i}$ respectively($i$=1,2).\\When $\beta$ is $\frac{1}{2}$ then this distance is called as Bhattacharya distance (BD)\cite{bhattacharyya1946measure}, which is given as 
\begin{equation}
J_B=\frac{1}{8}(\boldsymbol{\mu}_{2}-\boldsymbol{\mu}_{2})^{T}(\boldsymbol{\mu}_{2}-\boldsymbol{\mu}_{2})+\frac{1}{2}log\frac{(\frac{\left |\boldsymbol{\Sigma}_{1}+\boldsymbol{\Sigma}_{2}\right |}{2})}{\left | \boldsymbol{\Sigma}_{1} \right |^{\frac{1}{2}}\left |\boldsymbol{\Sigma}_{2}  \right |^{\frac{1}{2}}}
\end{equation}
\subsubsection{Ratio of Scatter Matrices}
In literature, the trace of ratio of scatter matrices (SR),is a measure of separability, as the trace of a scatter matrix is equal to the sum of the eigenvalues and therefore an indicator of the total variance in the data. How well features cluster around their class mean, as well as, how well they separate the class means. The scatter matrices, within-class scatter matrices,$\mathbf{S}_{w}$, and between class scatter matrices, $\mathbf{S}_{b}$, can be defined as
\begin{equation}
\mathbf{S}_{w}=\sum_{i=1}^{c}P_{i}E[(\mathbf{x}-\boldsymbol{\mu}_{i})^{T}(\mathbf{x}-\boldsymbol{\mu}_{i})]
\end{equation}
\begin{equation}
\mathbf{S}_{b}=\sum_{i=1}^{c}P_{i}(\boldsymbol{\mu}_{i}-\mathbf{\mu_{0}})^T(\boldsymbol{\mu}_{i}-\mathbf{\mu_{0}})
\end{equation}
where $\boldsymbol{\mu}_{i}$, $P_{i}$ and $\mu_{0}$  are mean vector of $i^{th}$ class data, prior probability of $i^{th}$ class data and global mean of data samples respectively.\\
From the definitions of scatter matrices, the criterion value which has to be maximized, is given as:
\begin{equation}
J_{SR}=\frac{trace(\mathbf{S}_{b})}{trace(\mathbf{S}_{w})}
\end{equation}
When intra cluster distance is very small and the inter cluster distance is very large $J_{SR}$ takes the high value. The main advantage of this criterion that it is not subject any external parameters and assumptions of any probability density function. Also the measure $J_{SR}$ under linear transformation has the advantage of being invariant under linear transformation.
\subsubsection{Linear Regression}
Linear regression is a statistical approach, which determines casual link of an independent variable upon a dependent variable. The class label of the data is recognized as the target dependent variable and the feature that affect the target is known as independent variable. There may be many features which can affect the class of the data, thus in such case multiple regression analysis would be more appropriate. A multiple regression model with $k$ independent features $\mathbf{f_1}, \mathbf{f_2}, \ldots, \mathbf{f_k}$ and a class variable $y$ can be written as \cite{park2007forward};
\begin{equation}
y_i=\beta _{0}+\beta _{1}X_{i1}+...+\beta_{k}X_{ik}+\zeta _{i}, i=1,2,...,n
\end{equation}
where $\beta_{0},\beta_{1},...,\beta_{k}$ is set of fixed values calculated by the class label $y$ and observed values of $\mathbf{X}$ and $\zeta _{i}$ is the error term. The sum of squared error (SSE) is given by
\begin{equation}
SSE=\sum_{i=1}^{n}(y_{i}-y_{i}^{p})^2
\end{equation}
where $y_{i}$ and $y_{i}^{p}$  are observed and predicated values respectively. The lower value of SSE depicts preferable regression model. The total sum of squares (SSTO) can be calculated as:
\begin{equation}
SSTO=\sum_{i=1}^{n}(y_{i}-\bar{y})^2
\end{equation}
where $\bar{y}$ is the mean value of $y_{i}, i=1,2,...,n$. The criterion function $J_{LR}$ is given as:
\begin{equation}
J_{LR}=R^2=1-\frac{SSE}{SSTO}
\end{equation}
The value of $J_{LR}$  lies between $0$ and $1$. The feature for which the value of $J_{LR}$ is higher is selected.
\subsubsection{minimum Redundancy-Maximum Relevance}
minimum-redundancy maximum-relevance (mRMR) \cite{park2007forward} is based on mutual information to discover a subset of features that have minimum redundancy among themselves and maximum relevance with the class labels. mRMR uses mutual information $I(\mathbf{f}_i,\mathbf{f}_l)$ as a measure of similarity between two feature vector $\mathbf{f}_i$  and $\mathbf{f}_l$, which is given as pursues:
\begin{equation}
I(\mathbf{f}_{i},\mathbf{f}_{l})=\sum _{k,l}p(\mathbf{f}_{k},\mathbf{f}_{l})\log(\frac{p(\mathbf{f}_{i},\mathbf{f}_{l})}{p(\mathbf{f}_{i})p(\mathbf{f}_{l})})
\end{equation}
where $p(\mathbf{f}_{i})$,$p(\mathbf{f}_{l})$ are the marginal probabilities of $k^{th}$ and $l^{th}$ features respectively and $p(\mathbf{f}_{i},\mathbf{f}_{l})$ is selected joint probability density. The relevance between the set of features S and the target class label vector $\mathbf{c}$, denoted by $REL$, is expressed as:
\begin{equation}
REL=\frac{1}{\left | S \right |}\sum_{\mathbf{f}_i\in S}I(\mathbf{f}_i,\mathbf{c})
\end{equation}
The average redundancy among features in the set $S$, denoted by $RED$, is defined as:
\begin{equation}
RED=\frac{1}{\left | S \right |^2}\sum_{\mathbf{f}_i,{\mathbf{f}_l}\in S}I(\mathbf{f}_i,\mathbf{f}_l)
\end{equation}
where $S$ denotes the subset of features and $\left | S \right |$ denotes the number of features in set $S$. Minimum redundancy and maximum relevance is measured by:
\begin{eqnarray}
\begin{split}
J_{MID}=max(f_{i})[REL-RED]=\\ max(f_{i})\left[\frac{1}{\left | S \right |}\sum_{\mathbf{f}_i\in S}I(\mathbf{f}_i,c)-\frac{1}{\left | S \right |^2}\sum_{\mathbf{f}_i,{\mathbf{f}_l}\in S}I(\mathbf{f}_i,\mathbf{f}_l)\right]
\end{split}
\end{eqnarray}
Clearly, the maximum values of $J_{MID}$ can be achieved with minimum redundancy among features and maximum relevance with target vector.
\section{Results and Discussion}\label{EX}
\subsection{Data}
\begin{figure}
\centering
\includegraphics[width=.35\textwidth]{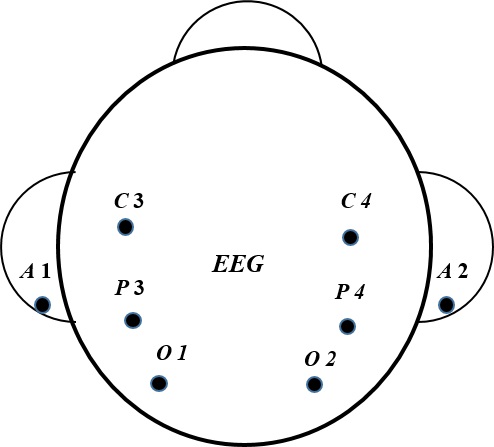}
\caption{Electrode placement of EEG recording adapted from\cite{palaniappan2002new}.}
\label{Fig-1}
\end{figure}

 For the simulation of our proposed framework, we have utilized a freely available Mental Task Classification data-set which has been used first time in the work of(Keirn and Aunon, 1990). Seven subjects (person) participated in the recording of this EEG dataset, but we did not utilize of Subject 4 due to incomplete information. In the baseline task (relax: B) each subject was instructed to carry out five different mental tasks ; the mental Letter Composing task (L); the Non trivial Mathematical task (M); the Visualizing Counting of numbers written on a blackboard task (C), and the Geometric Figure Rotation task (R). Each recording consists of the five trials of each of above said five mental tasks. EEG signals are recorded from C3, C4, P3, P4, O1 and O2 electrode position with A1 and A2 as the reference electrode as shown in Figure ~\ref{Fig-1}.
 Each trial is recorded for 10 seconds duration recorded with the sampling rate of 250 per second, which resulted in 2500 samples points per trial.

\begin{figure}
\centering
\includegraphics[width=.35\textwidth]{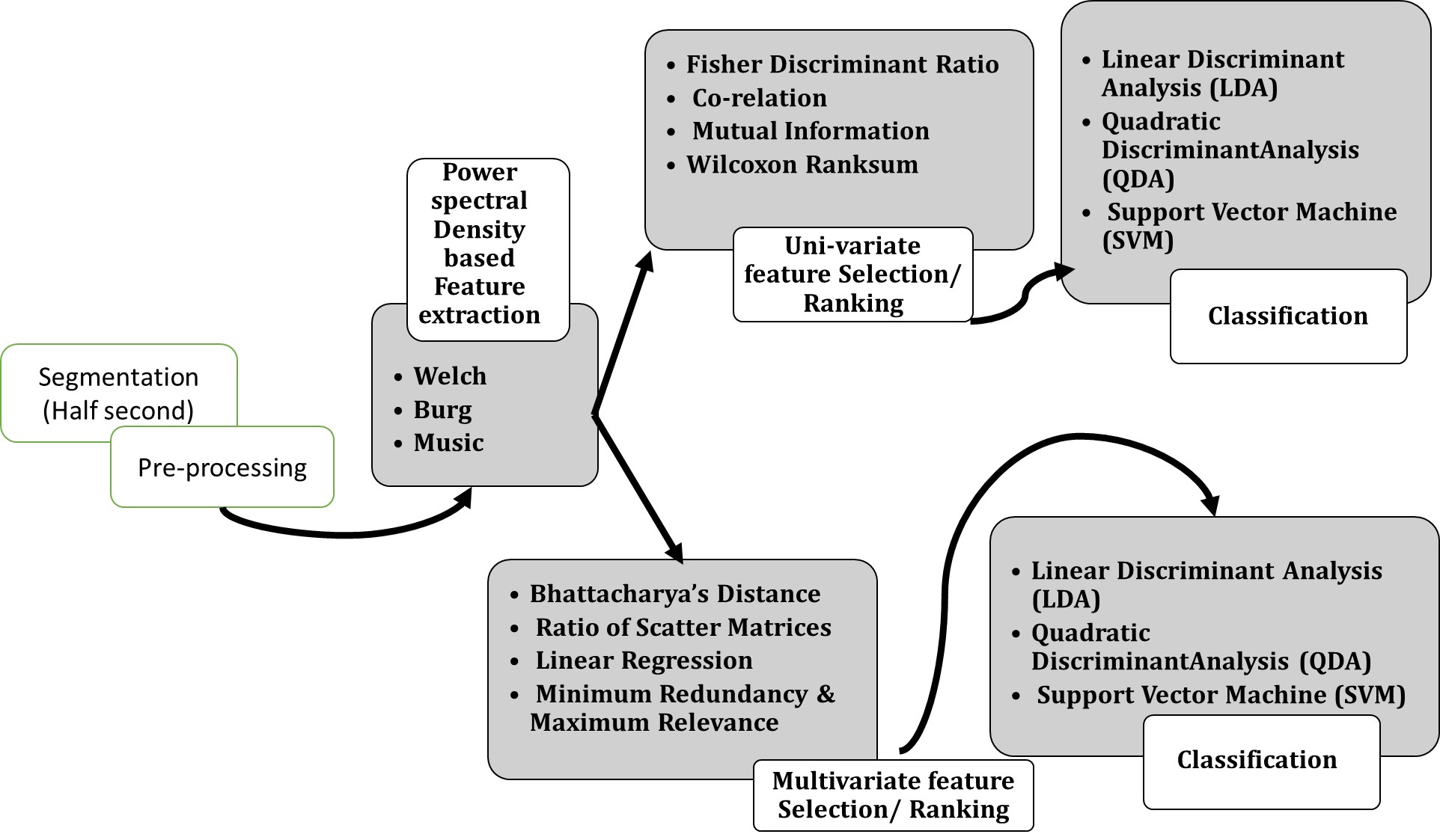}
\caption{Flow Diagram of the proposed approach for mental task classification.}
\label{Flow}
\end{figure}

Overall flow of the   proposed approach for mental task classification is shown in Figure  ~\ref{Flow}. The proposed approach consists four steps: segmentation feature extraction, feature selection and classification; to distinguish two different mental tasks. The main contribution of the work is employment of filter feature selection algorithm to enhance performance of learning algorithm for the classification of the mental tasks.
\subsection{Feature Formation}
\begin{figure*}
 \begin{minipage}{.5\textwidth}
  \includegraphics[width=0.28\textheight]{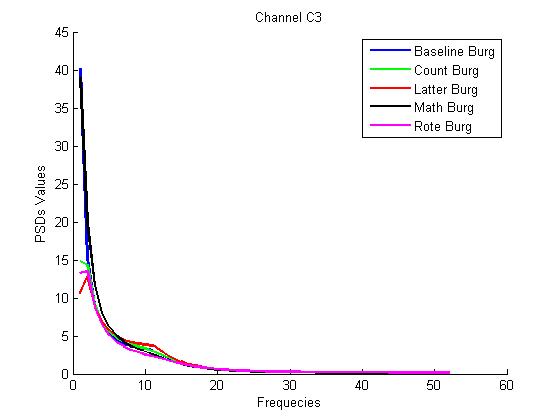}
 \end{minipage}
 \begin{minipage}{.5\textwidth}
  \includegraphics[width=0.28\textheight]{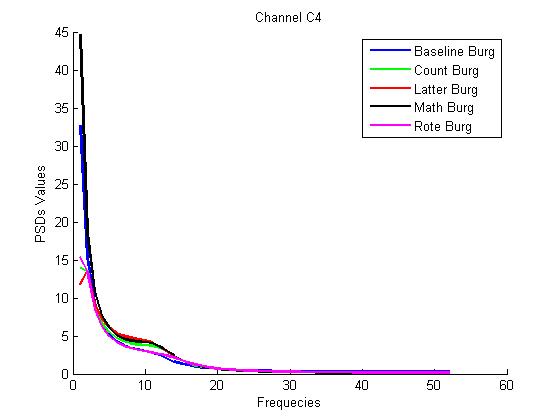}
 \end{minipage}
 \begin{minipage}{.5\textwidth}
  \includegraphics[width=0.28\textheight]{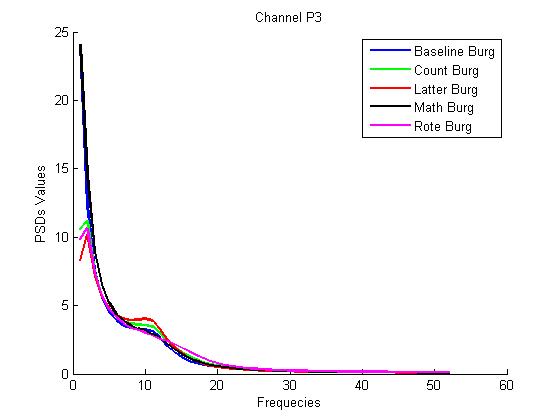}
 \end{minipage}
 \begin{minipage}{.5\textwidth}
  \includegraphics[width=0.28\textheight]{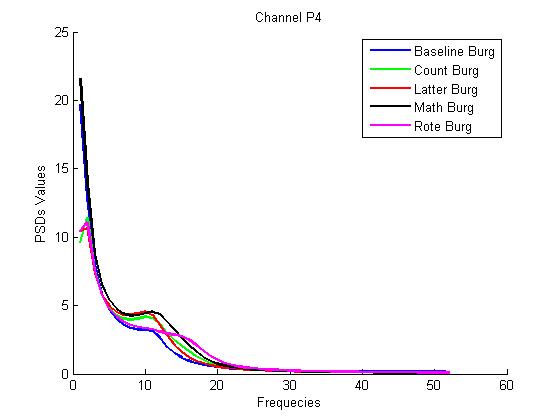}
 \end{minipage}
 \begin{minipage}{.5\textwidth}
  \includegraphics[width=0.28\textheight]{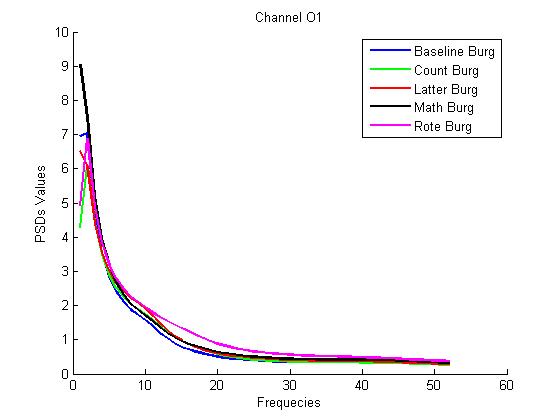}
 \end{minipage}
 \begin{minipage}{.5\textwidth}
  \includegraphics[width=0.28\textheight]{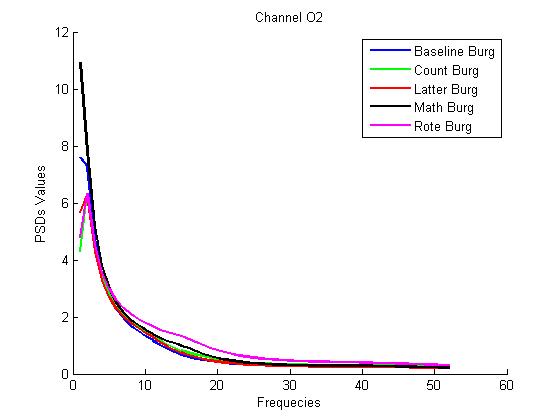}
 \end{minipage}
\caption{Comparison of features of different mental tasks using Burg method.}
\label{Fig-3}
\end{figure*}

For feature vector formulation, each trial data is pre-processed by decomposing into half-second segments, generating 20 fragments per trial for each subject. 
The extraction of features has been carried out from each signal using three different power spectral density approaches such as Welch, Burg, and MUSIC separately. A total of 52 PSD values are obtained from each channel. Combining PSD values of all six channels, each signal is represented regarding 312 PSD values. PSD values obtained for different tasks using Burg (parametric approach) for all six channels are shown in Figure ~\ref{Fig-3},
which shows that the extracting features from Burg PSD approach are effective in distinguishing different mental tasks. It can be also observed that PSD values at some frequency values differ considerably among different mental tasks (e.g. Frequency range of 6-9 Hz for channel C3, 6-13 Hz for channel C4, 6-13 Hz for channel P3, 6-16 Hz for channel P4, 6-9 Hz for channel O1 and 16-19 Hz for channel O2). This difference in PSD values can help in distinguishing different mental tasks. While PSD values at some frequency values take similar values (e.g. Frequency values above 15 Hz for C3, above 17 Hz for channel C4, above 13 Hz for channel O1, above 30 Hz for channel O2, above 20 Hz for channel P3 and above 22 Hz for channel P4) and cannot help in distinguishing different mental tasks.  Similar observations are also noted for Welch and MUSIC methods. This suggests that all features (PSD values) are not relevant for mental task classification.
\subsection{Application of Uni-variate Feature Selection}
\begin{figure*}
\centering
\includegraphics[width=.65\textwidth]{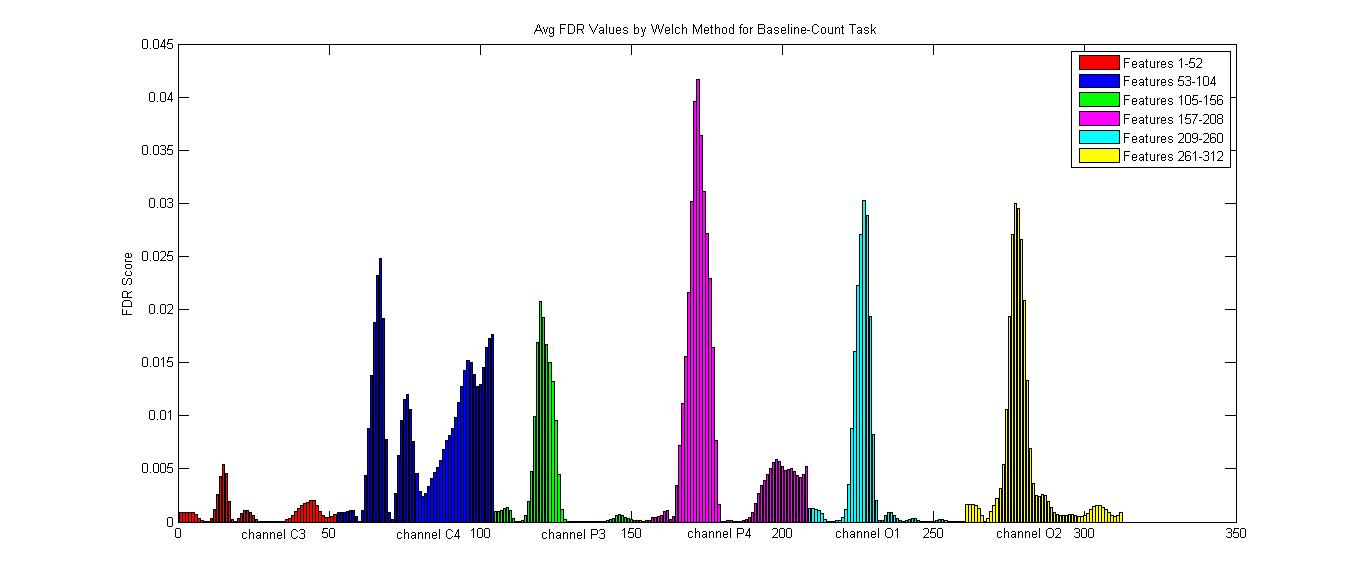}
\caption{Fisher Discriminant Ratio score for a pair of Baseline task and Count Task for features extracted using Welch.}
\label{Fig-4}
\end{figure*}

\begin{figure*}
\centering
\includegraphics[width=.65\textwidth]{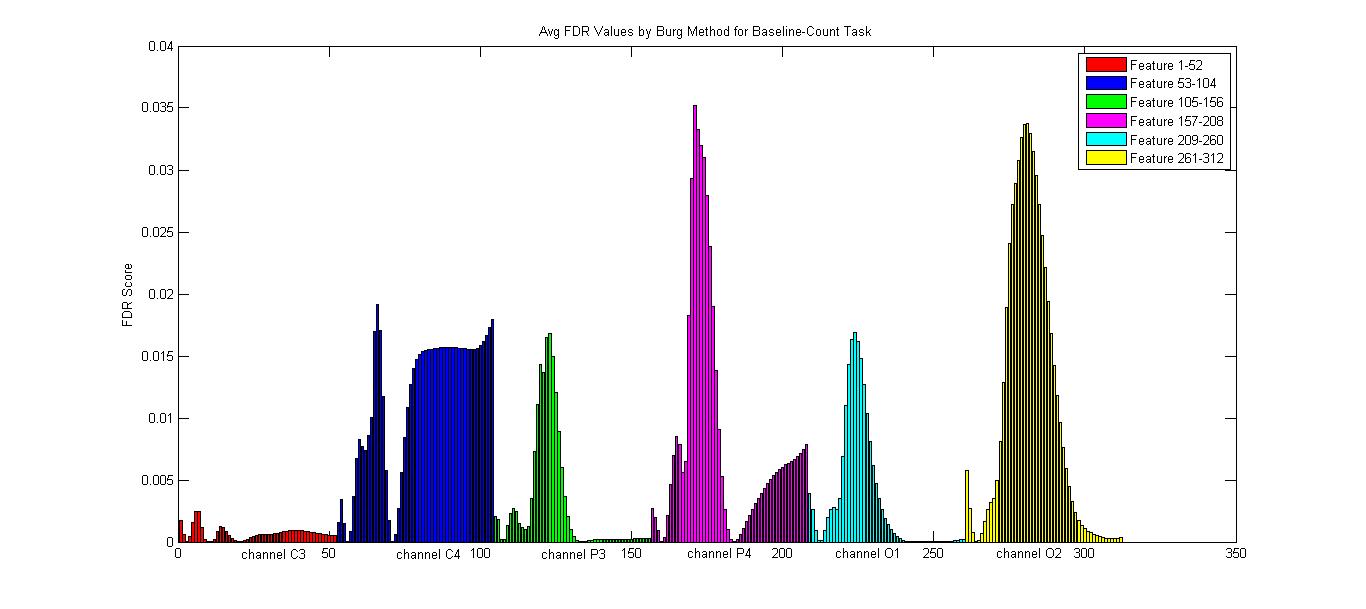}
\caption{Fisher Discriminant Ratio score for a pair of Baseline task and Count Task for features extracted using Burg.}
\label{Fig-5}
\end{figure*}

\begin{figure*}
\centering
\includegraphics[width=.65\textwidth]{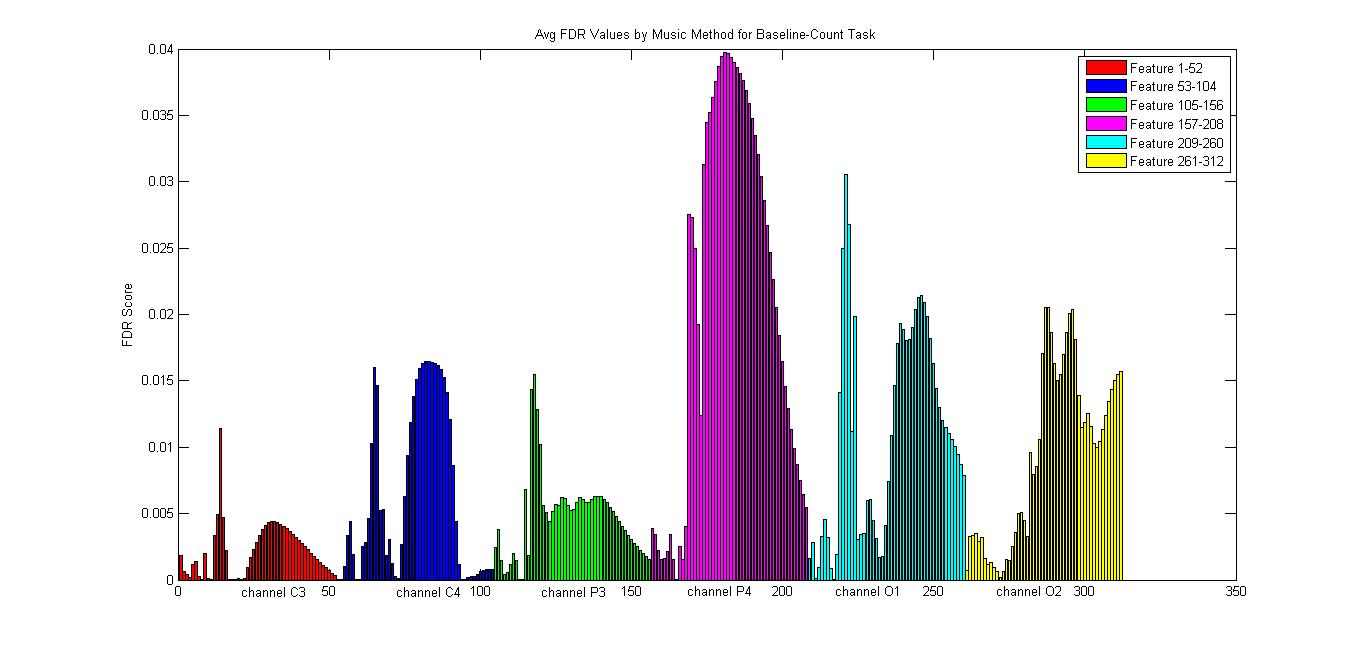}
\caption{Fisher Discriminant Ratio score for a pair of Baseline task and Count Task for features extracted using MUSIC.}
\label{Fig-6}
\end{figure*}

\begin{figure*}
\centering
\includegraphics[width=.65\textwidth]{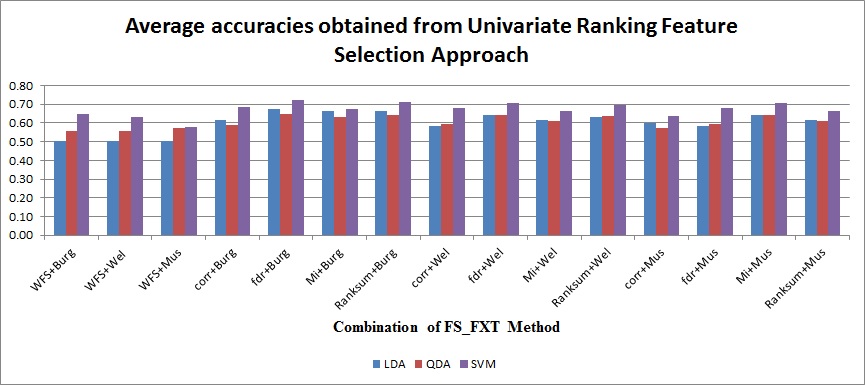}
\caption{ Comparison of different combination of univariate methods and PSD methods in terms of classification accuracy.}
\label{Fig-7}
\end{figure*}

To determine relevant features that can distinguish different mental tasks, four different univariate methods: Correlation (Cor), Fisher Discriminant Ratio (FDR), Mutual Information (MI) and Wilcoxon’s Rank Sum Test (Ranksum) are investigated in our experiment. FDR score corresponding to features obtained from each of the three PSD approaches to distinguish Baseline task from Count Task is shown from Figure ~\ref{Fig-4} to  Figure ~\ref{Fig-6}.
 It can be seen from these figures that FDR score corresponding to few features is very high and very less for others. This suggests that some features are more relevant than others. Similar observations are also noted for other univariate methods and other pairs of tasks.
For all univariate feature selection methods, the top 25 -ranked features are incrementally added to develop the decision model using forward feature selection approach. Comparison of different methods is reported in terms of maximum average classification accuracy for top features of 10 runs of 10 cross-validations. We have used three well-known classifiers: Linear Discriminant Analysis (LDA), Quadratic Discriminant Analysis (QDA) and Support Vector Machine (SVM) in our experiments. Figure ~\ref{Fig-7} shows a comparison of all combinations of three PSD approaches and four univariate methods with each of the three PSD approaches without any feature selection in terms of average classification accuracy (over six subjects for all combination of tasks).
We can observe the following from Figure ~\ref{Fig-7}
\begin{itemize}
\item    In general, the classification accuracy of all the three PSD approaches improves with the use of univariate feature selection method with all three classifiers. 
\item    Among all combinations of PSD approaches, univariate methods, and classifiers, the maximum classification accuracy is achieved with the combination of Burg, FDR, and SVM.
\item    Among four univariate feature selection methods, maximum classification accuracy is achieved with FDR.
\end{itemize}
\subsection{Application of Multivariate Feature Selection}
\begin{figure*}
\centering
\includegraphics[width=.65\textwidth]{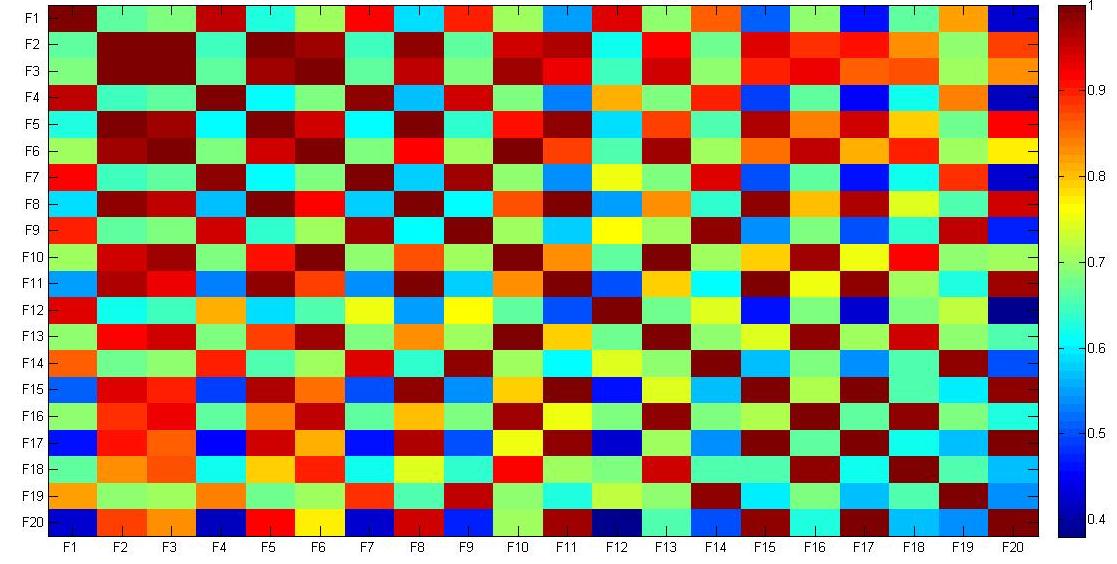}
\caption{ Colormap of Correlation values for top 20 PSD features obtained using combination of FDR .}
\label{Fig-8}
\end{figure*}

\begin{figure*}
\centering
\includegraphics[width=.65\textwidth]{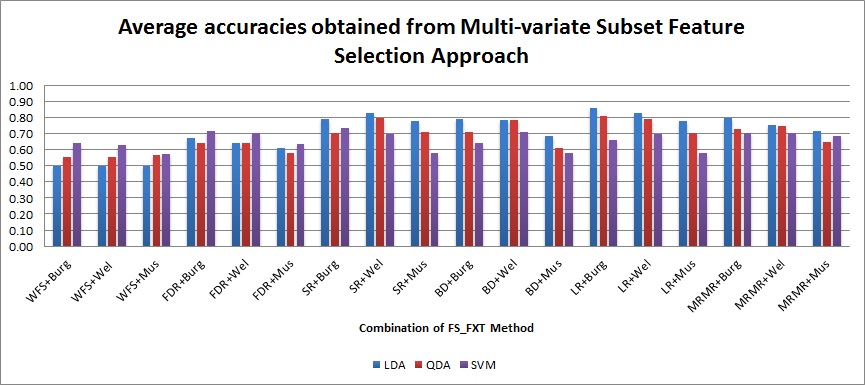}
\caption{ Comparison of all combinations of three PSD approaches and four multivariate methods with combination of PSD approaches and FDR in terms of average classification accuracy.}
\label{Fig-9}
\end{figure*}

Figure ~\ref{Fig-8} shows a color map of correlation values among top 20 relevant features obtained using the combination of FDR and Burg method to distinguish Baseline task from Count Task. 
It can be noted that some of the correlation values take a high value which depicts that such features are correlated (redundant) among themselves. Similar observations are also noted for other combinations of PSD approaches and univariate methods for another pair of tasks. This observation suggests the need to determine a subset of relevant and non-redundant features to further improve the performance of mental task classification.
For this, we used four well known multivariate methods: linear regression (LR), Bhattacharya distance (BD), Scatter Ratio (SR), Minimum Redundancy-Maximum Relevance (mRMR) to obtain minimal subset of non-redundant and relevant features using forward feature selection approach. Figure  ~\ref{Fig-9} shows a comparison of all combinations of three PSD approaches and four multivariate methods with the combination of PSD approaches and FDR (best performing univariate method) in terms of average classification accuracy.
 We can observe the following from Figure ~\ref{Fig-9}
\begin{itemize}
\item    Among all combinations of PSD approaches, multivariate feature selection methods and classifiers, the maximum classification accuracy is achieved with the combination of Burg, LR, and LDA.
\item    The performance of all combination of PSD approaches and multivariate methods is better in comparison to the combination of PSD approaches and FDR for LDA and QDA in terms of classification accuracy. 
\item    The performance of MUSIC is worst among three PSD approaches with univariate as well as multivariate feature selection methods.
\end{itemize}
\subsection{The Rankings of Respective Combinations of Feature Extraction and Selection Methods}
\begin{figure*}
\centering
\includegraphics[width=1\textwidth]{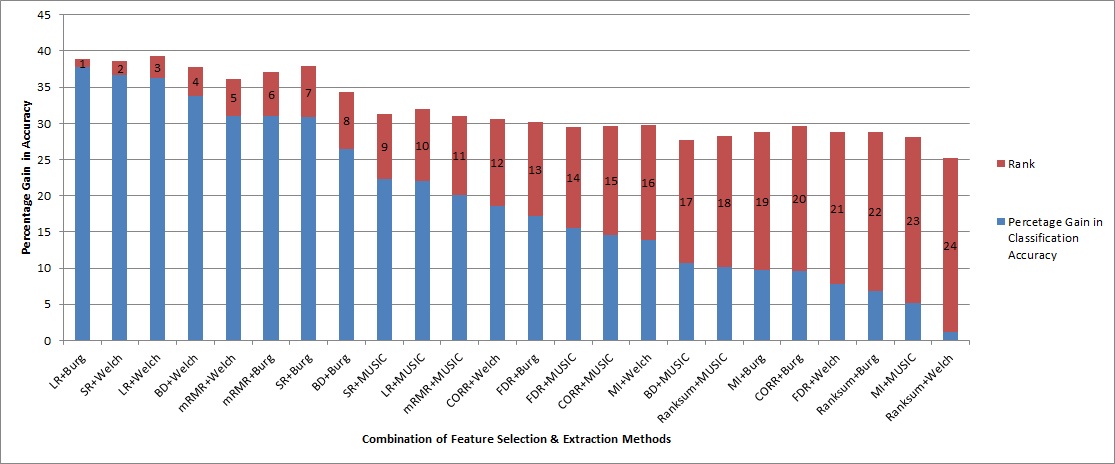}
\caption{ Ranking of different combinations of Feature Extraction and Selection methods}
\label{Fig-10}
\end{figure*}

To investigate the relational rank of both univariate and multivariate methods  feature selection techniques in combination with a feature extraction method, we have utilized the robust ranking approach \cite{adhikari2012performance}, on the ground of percentage gain in classification accuracy with respect to without applying any feature selection method \cite{gupta2017fuzzy}.\\
Figure ~\ref{Fig-10} shows twenty-four combinations of FS-FXT methods which are the feature selection and extraction methods. These methods are compared on the basis of percentage gain in accuracy of the different combination of selection and extraction methods and their corresponding ranks. From the Figure ~\ref{Fig-10}, we can observe that the combination of multivariate feature selection with all three feature extraction is ranked better in comparison to the combination of univariate feature selection and all three feature extraction methods except one combination (BD-MUSIC). Among all combination of selection and extraction methods, the combination of LR and Burg is best, whereas the team of MUSIC and Ranksum performs the worst.
\subsection{Friedman statistical test}
We have applied a non-parametric statistical test known as Friedman test in order to compare the statistically significant difference evolving in various combination of the feature selection and the PSD methods. 
From Table ~\ref{Table_4-12}, it can be noted that almost (11 out of 12) all combinations of multivariate feature selection with PSD methods obtained better rank than the combination of univariate feature selection method and PSD methods.
\begin{table}
  \centering
\caption{Average ranking using Friedman’s statistical test.}
    \begin{tabular}{|l|l|}
    \hline
    Different Combination & Ranking  \\
    \hline
    LR + Burg & \textbf{1.85}  \\
    \hline
    SR +  Welch & 2.2  \\
    \hline
    LR + Welch & 2.5  \\
    \hline
    BD + Welch & 3.84  \\
    \hline
    mRMR + Burg & 5.45  \\
    \hline
    SR +  Burg & 5.85  \\
    \hline
    mRMR + Welch & 6.3  \\
    \hline
    BD + Burg & 8.2  \\
    \hline
    SR + MUSIC & 10.25  \\
    \hline
    LR + MUSIC & 10.79  \\
    \hline
    mRMR + MUSIC & 11.2  \\
    \hline
    CORR + Welch & 11.6  \\
    \hline
    FDR + Burg & 12.65  \\
    \hline
    FDR + MUSIC & 14.15  \\
    \hline
    CORR + MUSIC & 15.3  \\
    \hline
    MI + Welch & 15.85  \\
    \hline
    CORR + Burg & 18.1  \\
    \hline
    RANKSUM + MUSIC & 18.7  \\
    \hline
    BD + MUSIC & 19  \\
    \hline
    MI + Burg & 19.15  \\
    \hline
    FDR + Welch & 20.35  \\
    \hline
    RANKSUM + Burg & 20.9  \\
    \hline
    MI + MUSIC & 22.15  \\
    \hline
    RANKSUM + Welch & 23.649  \\
    \hline
    \end{tabular}%
  \label{Table_4-12}%
\end{table}%

The SEL-EXT pair performance is also examined with respect to a control method, i.e., the one that emerges with the lowest rank (combination of LR and Burg). 
In the comparison of the control method with other 23 combinations of feature selection and feature extraction method, adjusted p-values \cite{derrac2011practical} we computed in order to take into account the error accumulated and provide the correct correlation. A set of post-hoc procedures to provide correct correlation is defined in the literature. The adjusted p-values in the method are computed in order to find whether the control method shows any statistical difference when compared with the remaining methods. For pair-wise comparisons, the widely used post hoc methods to obtain adjusted p-values are \cite{derrac2011practical}: Holm, Hochberg and Hommel procedures. Table ~\ref{Table_4-13} illustrates adjusted p-values for the Hommel procedure.
\begin{table}
  \centering
 \caption{Adjusted p-values for the Hommel procedure.}
    \begin{tabular}{|l|l|l|}
    \hline
   Different Combination & unadjusted p & p Homm  \\
    \hline
    Ranksum + Welch  & \textbf{5.43E-12} & \textbf{1.25E-10}  \\
    \hline
    MI + MUSIC  & \textbf{1.37E-10} & \textbf{3.01E-09}  \\
    \hline
    Ranksum + Burg  & \textbf{1.70E-09} & \textbf{3.57E-08}  \\
    \hline
    FDR + Welch & \textbf{4.91E-09} & \textbf{9.82E-08}  \\
    \hline
    MI + Burg  & \textbf{4.48E-08} & \textbf{8.07E-07}  \\
    \hline
    BD + MUSIC & \textbf{5.85E-08} & \textbf{9.95E-07}  \\
    \hline
    Ranksum + MUSIC & \textbf{9.91E-08} & \textbf{1.68E-06}  \\
    \hline
    CORR + Burg  & \textbf{2.77E-07} & \textbf{4.43E-06}  \\
    \hline
    MI + Welch  & \textbf{9.55E-06} & \textbf{1.43E-04}  \\
    \hline
    CORR + MUSIC & \textbf{2.11E-05} & \textbf{2.95E-04}  \\
    \hline
    FDR + MUSIC & \textbf{1.00E-04} & \textbf{0.001305}  \\
    \hline
    FDR + Burg & \textbf{6.37E-04} & \textbf{0.007647}  \\
    \hline
    CORR + Welch  & \textbf{0.0020477} & \textbf{0.020477}  \\
    \hline
    mRMR + MUSIC & \textbf{0.0031092} & \textbf{0.027983}  \\
    \hline
    LR + MUSIC & \textbf{0.0046513} & \textbf{0.037211}  \\
    \hline
    SR + MUSIC  & \textbf{0.0079} & 0.0632  \\
    \hline
    BD + Burg  & \textbf{0.0446384} & 0.312469  \\
    \hline
    mRMR + Welch  & 0.1593641 & 0.637456  \\
    \hline
    SR +  Burg & 0.2059032 & 0.823613  \\
    \hline
    mRMR + Burg & 0.2549452 & 0.91187  \\
    \hline
    BD + Welch  & 0.5270893 & 0.91187  \\
    \hline
    LR + Welch & 0.837144 & 0.91187  \\
    \hline
    SR +  Welch & 0.9118703 & 0.91187  \\
    \hline
    \end{tabular}%
  \label{Table_4-13}%
\end{table}%

The values in Table ~\ref{Table_4-13} represents the p-value when the pair-wise comparison with control method(Burg+LR) is conducted. The bold values suggest the significant difference observed from the control method (Burg+LR) with the combinations at the significance level of $0.05$. This demonstrates that combination of Burg with LR performs significantly better than all combinations of univariate method and feature extraction methods. It also performs significantly better than few combinations of multivariate method and feature extraction method.
\section{Conclusion} \label{CON}
In this paper, we have examined the performance of the combination of three different PSD approaches, with four well-known uni-variates as well as four very popular multi-variates, filter feature selection methods. The experimental findings demonstrate that the multivariate feature selection algorithms endue more distinguishable feature set for the mental task classification, compared with univariate feature selection approach. The outcome determined features for the mental task classification by a minimal subset of relevant and non-redundant features. Experimental results demonstrate significant improvement in classification accuracy utilizing the selected feature selection methods. It is observed that the performance of multivariate filter feature selection methods is, in general, better than univariate filter feature selection methods. The combination of Burg method, LR and Linear Discriminant Analysis(LDA) achieved maximum classification accuracy among all other combinations.

 Statistical tests also endorsed that the performance of the combination of Burg and the linear regression is notably different from the majority of the combinations. It has also been observed that for mental task classification multivariate feature selection approach works better than univariate feature selection approach in most of the cases with the conjunction of power spectral density approach. 

 In the future, we would like to extract spectral density of different brain frequency separately. Since the comparisons and investigations have been done on binary mental task classification, we would, therefore, like to extend this approach for multi-class mental tasks classification. 
\section*{Acknowledgements}
The first author would like to express his gratitude to the Council of Scientific \& Industrial Research (CSIR), India, and acknowledge the financial support for the research work.
\bibliographystyle{IEEEtran}
\bibliography{mycite}
\end{document}